\documentclass[12pt]{article}

\usepackage{amsmath,amssymb}

\usepackage[dvips]{graphicx}

\newcommand{\be}{\begin{equation}}
\newcommand{\ee}{\end{equation} }
\newcommand{\bea}{\begin{eqnarray}}
\newcommand{\eea}{\end{eqnarray} }
\newcommand{\ba}{\begin{array}}
\newcommand{\ea}{\end{array}}
\newcommand\dis{\displaystyle}
\newcommand{\nn}{\nonumber}
\newcommand{\bOmega}{\overline{\Omega}}

\newcommand{\unit}{\mathbf{1}}                              
\newcommand{\eps}{\varepsilon}                              
\newcommand{\grSU}{\mathrm{SU}}                             
\newcommand{\grSO}{\mathrm{SO}}                             

\newcommand{\half}{{{\textstyle\frac{1}{2}}}}
\newcommand{\quarter}{{{\textstyle\frac{1}{4}}}}

\newcommand{\Zb}{{\bar Z}}

\newcommand\tr{{\rm tr}}
\newcommand\Tr{{\rm Tr}}

\newcommand\cL{{\cal L}}
\newcommand\cM{{\cal M}}
\newcommand\cN{{\cal N}}

\newcommand\cW{{\cal W}}

\newcommand\rmd{{\rm d}}

\newcommand{\eqn}[1]{(\ref{#1})}
\newcommand{\PB}[2]{\{ #1 , #2\}}

\begin{document}

\thispagestyle{empty}
\begin{flushright}
{\sc\small HU-EP-08/08}
\end{flushright}
\vspace{0.5cm}
\setcounter{footnote}{0}
\begin{center}
{\Large{\bf \mathversion{bold} $\cM$-theory on pp-waves with a
holomorphic superpotential and its membrane and matrix  descriptions
\mathversion{normal}
\par}
    }\vspace{20mm}
{\sc \large Jongwook Kim${}^{1}$, Nakwoo Kim${}^{2}$,  Jeong-Hyuck Park${}^{1}$ and Jan Plefka${}^{3}$}\\[7mm]
{\it ${}^1$Department of Physics, Sogang University,  Seoul 121-742, Korea\\[5mm]
${}^2$Department of Physics and Research Institute of Basic Science,
Kyung Hee University, Seoul 130-701, Korea \\[5mm]
${}^3$Institut f\"ur Physik,
Humboldt-Universit\"at zu Berlin,
Newtonstra{\ss}e 15, D-12489 Berlin, Germany
}\\ [5mm]
{\tt jongwook@sogang.ac.kr, park@sogang.ac.kr, nkim@khu.ac.kr, plefka@physik.hu-berlin.de }\\[15mm]

{\sc Abstract}\\[2mm]
\end{center}
We study a new class of inhomogeneous pp-wave solutions with 8
unbroken supersymmetries in $D=11$ supergravity. The 9 dimensional
transverse space is Euclidean and split into 3 and 6 dimensional
subspaces. The solutions have non-constant gauge flux, which are
described in terms of an arbitrary holomorphic function of the
complexified 6 dimensional space.  The supermembrane and matrix theory
descriptions are also provided and we identify the relevant
supersymmetry transformation rules. The action also arises through a
dimensional reduction of ${\cal N}=1, D=4$ supersymmetric Yang-Mills
theory coupled to 3 gauge adjoint and chiral multiplets, whose
interactions are determined by the holomorphic function of the
supergravity solution now constituting the superpotential.

\hspace*{-.4cm}

\newpage

\setcounter{page}{1}


\tableofcontents 

\section{Introduction}
Over the years it has become more evident that string theory, as a
candidate of quantum gravity, and Yang-Mills theory are dual to each
other \cite{Maldacena:1997re}. One important line of progress has been
made around the Matrix theory conjecture \cite{Banks:1996vh}, which
suggests that  M-theory, the quantum completion of 11 dimensional
supergravity theory in Minkowski background, is reduced to
supersymmetric Yang-Mills quantum mechanics with $SU(N)$ gauge group
in the large $N$ limit, when viewed in light-cone frame. There are two
seemingly different ways to justify the Matrix theory conjecture. One
is as a discretized supermembrane action \cite{de Wit:1988ig}, and the
other is as the D0-brane dynamics which is believed to give a partonic
description of M-theory when quantized along a light-like direction
\cite{Banks:1996vh}.

It is certainly desirable to extend the Matrix theory conjecture to
more general backgrounds with less supersymmetry and smaller isometry
groups, see {\it e.g.} \cite{Kim:1997uv,Kim:1997gh}. A natural way to
explore is to turn on the gauge flux.  Indeed, when one considers
the maximally supersymmetric plane-wave solution
\cite{KowalskiGlikman:1984wv}, it is again possible to find the
supermembrane action in light-cone gauge and
the resulting Yang-Mills quantum mechanics is conjectured to give the
corresponding Matrix theory description \cite{Berenstein:2002jq,Dasgupta:2002hx}. This
particular matrix model is usually called the BMN (Berenstein,
Maldacena, and Nastase) matrix model, and thanks to the mass parameter
set by the non-vanishing flux, one can perturbatively compute the
energy spectrum \cite{Dasgupta:2002hx,Kim:2002if}, unlike the original Matrix
model in flat background. The existence of protected supermultiplets
\cite{Kim:2002if,Dasgupta:2002ru,Kim:2002zg} turns out to be essential
to verify the duality at the non-perturbative level, {\it e.g.} the dual modes
of transverse M5-branes in the matrix model \cite{Maldacena:2002rb}.
It is also noteworthy that the mass parameter of the Matrix theory can be traced
back to the radius of $S^3$, when one puts the superconformal ${\cal N}=4, D=4$
Yang-Mills theory on $\mathbb{R}\times S^3$ for dimensional reduction
\cite{Kim:2003rza}.

It is straightforward to consider similar plane-wave solutions of $D=11$ supergravity
with
constant flux and less supersymmetries
\cite{Cvetic:2002si,Gauntlett:2002cs,Lee:2002vx}. One notable feature of such
solutions is the so-called {\it supernumerary} supersymmetries, which mean that
they preserve between 16 to 32 supersymmetries. It is possible for
some of such backgrounds to identify the string/M-theory origin, for
instance as intersecting M-brane configurations \cite{Kim:2002cr}.

In this paper we attempt a further generalization and consider
pp-waves with {\it non-constant} gauge fluxes. More specifically, we divide
the 9 dimensional transverse space into a real 3 dimensional space $\mathbb{R}^3$ and a
complex 3 dimensional subspace $\mathbb{C}^3$, and allow the configurations to
depend only on the coordinates of $\mathbb{C}^3$ through a holomorphic
function, which we call a superpotential. As a result the metric tensors will have
$SO(3)\times U(1) \times \mathbb{R}$ isometry,
where $U(1)$ is the remaining invariance of the complex 3 dimensional space guaranteed
by holomorphicity, and $\mathbb{R}$ denotes the null Killing vector.
However, $U(1)$ is generically broken for the entire solution when we also take the flux
into account.

Inhomogeneous pp-waves with non-constant flux have been considered by several
authors in similar settings. For 10 dimensional IIB supergravity, pp-waves on
special holonomy manifolds are studied and it is shown that the Ramond-Ramond 5-form
induces a superpotential on the light cone worldsheet Lagrangian \cite{Maldacena:2002fy},
while the 3-forms are responsible for Killing vector potentials \cite{Kim:2002gi}.
More recently, inspired by \cite{Lunin:2005jy}, supersymmetric Matrix models with
so-called $\beta$-deformation are studied \cite{shimada}, and it is illustrated
that the deformation superpotential of the discretized supermembrane action
is given by inhomogeneous background fluxes which have a
linear dependence on transverse coordinates. It is this discovery which
motivated our research on pp-waves with a generic superpotential reported in this paper.

We establish the light-cone supermembrane action in our new
inhomogeneous pp-wave configurations and write down the relevant
matrix model action, which turns into the supermembrane action in the
continuum large $N$ limit. Because this matrix model has a generic superpotential,
whose arguments are promoted to matrices, we encounter the usual
matrix ordering ambiguity problem. We give an exposition on how this ambiguity
is fixed by supersymmetry and the requirement to express the superpotential as
a gauge singlet, i.e.~a single or multi-trace operator.

The symmetry of our solutions and the existence of a holomorphic function suggests
that these models should be naturally related to ${\cal N}=1, D=4$ super Yang-Mills theory
with 3 chiral multiplets in the adjoint representation.
It is verified explicitly through decomposition of the
fermionic as well as the bosonic fields, and we identify the total superpotential of
the Yang-Mills quantum mechanics.

This paper is organised as follows. In Sec. 2, we present the pp-wave
solutions of 11 dimensional supergravity we will be dealing with in
this paper. It is also shown that these backgrounds in general
allow 8 nontrivial solutions to the Killing spinor equation. In
Sec. 3, we consider the light-cone action of supermembranes in the
pp-wave background, and show how it is reduced to a gauge theory of
area-preserving diffeomorphisms. In Sec. 4 we provide the Yang-Mills
quantum mechanics action which is obtained via the usual
discretization method of replacing the Poisson brackets with
commutators \cite{HoppePhD,de Wit:1988ig}. In Sec. 5, we establish
that our supermembrane/matrix actions can be also expressed as ${\cal
N}=1$ Yang-Mills theory with 3 interacting chiral multiplets, and identify
the total superpotential. In Sec. 6 we conclude with brief discussions.

\section{A class of supersymmetric pp-wave solutions in $D=11$ supergravity}
Let us start by presenting the supergravity solutions we will
study in this paper. Readers are referred to, for instance \cite{Gauntlett:2002cs},
for conventions of $D=11$ supergravity.

Most generally, by a pp-wave in 11 dimensional supergravity we mean the following type of configurations
\bea\label{sugraansatz}
ds^2 &=& 2 dx^- dx^+ + H(x^+ , x^M ) (dx^+)^2 + \sum_{M=1}^9 (dx^M)^2 ,
\\
F^{(4)} &=& dx^+ \wedge \phi (x^+, x^M) .
\eea
The above ansatz is greatly simplifying and one can easily verify that the only nontrivial component of the Einstein equation gives
\be
\nabla^2 H =  -\frac{1}{6} \phi_{MNP} \phi^{MNP} ,
\label{sole}
\ee
where the Laplacian is taken in the 9 dimensional transverse space.
One of course has to consider the flux equation of motion and the Bianchi identity for
$F^{(4)}$,  so we demand $\phi$ is harmonic:
\be
d\phi = 0 , \quad d (*_9 \phi ) = 0 .
\ee

In this paper we are interested in a rather special subclass of the general
pp-waves given above.
We first divide the 9 dimensional space into 3 dimensional and 6 dimensional subspaces,
and assume $H, \phi $ can depend only on the 6 dimensional coordinates.
We will find it convenient to employ complex coordinates for the 6 dimensional space.
Let us call them $z_a, (a=1,2,3)$, and $\bar{z}_{\bar{a}}$ are the complex conjugates.
Now the 9 dimensional part of the metric is written as
\be
ds^2_9  = \sum_{i=1}^{3} (dx^i)^2 + 2 \sum_{a=1}^3 dz^a d\bar{z}^{\bar{a}} ,
\ee
and accordingly $\nabla^2= 2 \sum_{a=1}^3 \partial_a \bar{\partial}_{\bar{a}} $.

The upshot is that we can obtain a large class of solutions which are reminiscent of
${\cal N}=1, D=4$ supersymmetric field theory. Our construction is as follows.
Firstly, we choose $\phi $ as a primitive $(2,1)$ form plus its complex conjugate,
in the space spanned by $z_a$.
Componentwise,  one writes
\bea
\phi_{\bar{a}bc} &=& \partial_{\bar{a}}\partial_{\bar{d}} \overline{W}
\epsilon^{\bar{d}}_{\;\; bc} ,
\label{flux}
\eea
and in the same way for the complex conjugate, $\phi_{a\bar{b}\bar{c}}$.
Note that $W$ is a holomorphic function of $z^a$, and we take the
convention $\epsilon_{123}=\epsilon_{\bar{1}\bar{2}\bar{3}}=1$
for the totally antisymmetric tensor. Primitivity means that a symplectic trace
of $\phi$ is zero, {\it i.e.} $\phi_{\bar{a}ab}=0$, implying $\phi$ is imaginary self-dual
in 6 dimensions.

It can be easily confirmed that $\phi$, as given above, is in fact closed and co-closed.
One easily integrates Eq.(\ref{flux}) and the 2-form potential $\psi$, with $d\psi=\phi$,
is given in terms of a $(2,0)$ form,
\be
\psi_{ab} = \epsilon^{\bar{c}}_{\;\;ab} \partial_{\bar{c}} \overline{W} .
\ee

Now we only need to check the equation Eq.(\ref{sole}) with an appropriate choice of $H$.
It is easily seen that Eq.(\ref{sole}) is indeed satisfied with
\be
H = - | \partial W |^2 .
\ee

Having established that the configuration indeed satisfies the equations of motion,
let us now consider the Killing spinor equations.
For 11 dimensional pp-waves with constant flux, the Killing spinor equations have
been studied in detail in Ref. \cite{Cvetic:2002si,Gauntlett:2002cs,Lee:2002vx}. We closely
follow the convention and the analysis of \cite{Gauntlett:2002cs}, which is
repeated here to some extent for self-sufficiency.

For the pp-wave metric, it is convenient to choose the following frame
\bea
e^+ &=& dx^+ ,
\\
e^- &=& dx^- + \frac{H}{2} dx^+ ,
\\
e^M &=& dx^M ,
\eea
then the only nonvanishing components of the spin connection are
\be
\omega^{-M} = \frac{1}{2} \partial_M H dx^+ .
\ee

\relax From the $D=11$ supersymmetry transformation rule, the invariance of the
gravitino requires $\nabla_\mu \epsilon = \Omega_\mu \epsilon$, with
\be
\Omega_\mu = \frac{1}{288} ( \gamma_{\mu}^{\;\;\;\;\nu\rho\sigma\tau} -
8 \delta^\nu_\mu \gamma^{\rho\sigma\tau} ) F^{(4)}_{\nu\rho\sigma\tau}  .
\ee
\relax From the pp-wave ansatz $\Omega_\mu$ is reduced to
\bea
\Omega_+ &=& -\frac{1}{12} \Theta ( \gamma_{-} \gamma_{+} + 1 ) ,
\\
\Omega_- &=& 0 ,
\\
\Omega_M &=& \frac{1}{24} ( 3 \Theta \gamma_M + \gamma_M \Theta ) \gamma_- ,
\eea
where $\Theta = \frac{1}{6}\phi_{MNP} \gamma^{MNP}$.

At this stage, it is convenient to introduce two $SO(9)$ spinors, $\epsilon_{\pm}$,
to describe the 11 dimensional Majorana spinor $\epsilon$.
We use the basis where $\gamma_\mu$ are  $32\times 32$ matrices
and given as
\be
\gamma_+ = \begin{pmatrix}0 & 1 \\ 0 & 0 \end{pmatrix} ,
\quad \quad
\gamma_- = \begin{pmatrix}0 & 0 \\ 1 & 0 \end{pmatrix} ,
\quad\quad
\gamma_M = \begin{pmatrix}\Gamma_M & 0 \\ 0 & -\Gamma_M \end{pmatrix} ,
\ee
where $M=1,\ldots, 9$ and $\Gamma_M$ are $SO(9)$ gamma matrices.
We can in fact take $\Gamma_M$ to be all real and symmetric.
 Now, if we accordingly decompose $\epsilon$ as
\be
\epsilon = \begin{pmatrix}
\epsilon_+ \\ \epsilon_-
\end{pmatrix} ,
\ee
we have the following equations for $\epsilon_\pm$.
\be
\ba{ll}
\textstyle{\partial_+ \epsilon_+} = -\textstyle{\frac{1}{12}} \Theta \epsilon_+\,,~~~&~~~
\textstyle{\partial_+ \epsilon_-} =-\textstyle{\frac{\sqrt{2}}{4}} \partial \hspace{-6pt}
/ H \epsilon_+ + \textstyle{\frac{1}{12}}\Theta \epsilon_- \,,\\
{}&{}\\
\partial_- \epsilon_+ = 0 \,,~~~&~~~
\partial_- \epsilon_- = 0 \,,\\
{}&{}\\
\partial_M \epsilon_+ = 0 \,,~~~&~~~
\partial_M \epsilon_- = \textstyle{\frac{\sqrt{2}}{24}}
( 3\Theta \Gamma_M + \Gamma_M \Theta ) \epsilon_+ \,.
\ea
\ee
With a slight abuse of the notation, we now re-defined $\Theta=\frac{1}{6}\Phi_{MNP}\Gamma^{MNP}$.
We see that, in general we can first solve the equations for $\epsilon_+$,
then plug it into the equations for $\epsilon_-$.
A simple type of solutions, which are sometimes called {\it kinematic} supersymmetries,
are given as follows: We set $\epsilon_+=0$, and demand $\epsilon_-$ is also constant
and annihilated by $\Theta$. Since our 3-form field $\phi$ is $(2,1)$ and primitive,
$\Theta$ annihilates any $SU(3)$ singlet spinor, which satisfies the following
projection rules
\be
\label{singlet}
\Gamma_{12}\epsilon
=\Gamma_{34}\epsilon
=\Gamma_{56}\epsilon .
\ee
Let us denote hereafter, a constant and Majorana spinor of SO(9) satisfying Eq.(\ref{singlet}) as
$\epsilon^{(0)}$. It is now obvious that
\be
\epsilon_+ = 0 , \quad \epsilon_- = \epsilon^{(0)} ,
\ee
provides 4 linearly independent solutions of the Killing equation.

Now let us verify that our background configuration in fact allows 4 more supersymmetries
with $\epsilon_+ \neq 0$. This is sometimes called {\it dynamical} supersymmetries, and are
responsible for the supersymmetry of the supermembrane action or the associated super
Yang-Mills action which will be derived in the remainder of this paper.  We already know
that the equation for $\epsilon_+$ can be solved by any constant spinor if it is an $SU(3)$ singlet.
So we first set $\epsilon_+ = \epsilon^{(0)}$. After a little computation, one can verify that
if we set
\be
\epsilon_- = -\frac{\sqrt{2}}{8}\partial_a W \epsilon^{abc} \Gamma_{bc}  \epsilon^{(0)} + c.c.
\ee
then the equations for $\epsilon_-$ are identically satisfied.
Note that our Killing spinor solutions, kinematical and dynamical altogether,
have no dependence on $x^+$.

\section{The light-cone supermembrane description}
We now wish to derive the light-cone gauge fixed action for a supermembrane propagating in the
above pp-wave background. The structure of the (super)-membrane action in curved backgrounds
has been analyzed in a number of works, including \cite{deWit:1998tk,Kim:2006wg}.
We here briefly repeat this construction in an
economic first order formalism.

\subsection{First order formalism}

In order to fix the light-cone gauge it is advantageous to bring the Polyakov formulation of
the membrane into a first order formulation. The bosonic membrane propagating in a general
background geometry $G_{\mu\nu}(X)$ and 3-form potential $C_{\mu\nu\rho}(X)$ with
 the membrane embedding coordinates $X^\mu=X^\mu(\tau,\sigma_1,\sigma_2)$ is given by
\begin{align}
S= -\frac{T}{2}\, &\int d^3\xi \, \Bigl ( \gamma^{\alpha\beta}\, \partial_\alpha X^\mu\, \partial_\beta X^\nu\,
G_{\mu\nu}(X) - \sqrt{-h}\, \Bigr ) \nonumber\\
& + \frac{\kappa}{3!}\, \int d^3\xi \, \epsilon^{\alpha\beta\gamma}\,
\partial_\alpha X^\mu\, \partial_\beta X^\nu\,
\partial_\gamma X^\rho\, C_{\mu\nu\rho}(X) ,
\end{align}
where $h_{\alpha\beta}$ is the world-volume metric and we have defined
$\gamma^{\alpha\beta}= \sqrt{-h}\, h^{\alpha\beta}$ with $\det \gamma = - \sqrt{-h}$.
$T$ is the membrane tension and $\kappa=\pm 1$.
This `Polyakov' form of the action
is equivalent to the first order formulation
\begin{align}
S' = \int d^3\xi \, &\Bigr [ P_\mu\, \dot X^\mu + \frac{1}{2T\gamma^{00}}\, \Bigl \{ \,
P_\mu\, P_\nu\, G^{\mu\nu}(X) + T^2\, \gamma^{0r}\, \partial_rX^\mu\, \gamma^{0s}\, \partial_sX^\nu\,
G_{\mu\nu} \nonumber\\ &
- \kappa \, P_\mu\, G^{\mu\nu}\, C_{\nu\rho\kappa}\, \epsilon^{rs}\, \partial_rX^\rho\,
\partial_sX^\kappa \, \Bigr \}
 -\frac{T}{2}\, \Bigl ( \, \gamma^{rs}\, \partial_rX^\mu\, \partial_sX^\nu\, G_{\mu\nu}(X) \nonumber\\ &
+\det \gamma \Bigr )
+ \frac{\gamma^{0r}}{\gamma^{00}}\, P_\mu\, \partial_rX^\mu\, \Bigl ] .
\label{FOFORM1}
\end{align}
Here $r,s=1,2$ denote the space-like directions on the membrane world-volume.
One checks that plugging back into $S'$ the solution of the algebraic field equations for $P_\mu$ yields
$S$. The equations of motion for the non-dynamical $\gamma^{\alpha\beta}$ give rise to the
constraints of the theory. We now proceed by choosing the gauge condition
$\gamma^{0r}=0$ which turns its associated
constraint equation into
\begin{equation}
P_\mu\, \partial_rX^\mu = 0 \, .\label{Constraint}
\end{equation}
This is the analogue of the level matching condition in string theory. Furthermore the equation of motion
for $\gamma^{rs}$ can be solved to give
\begin{equation}
\gamma^{rs}= \frac{1}{\gamma^{00}}\, \left ( \begin{matrix} -\partial_2X \cdot \partial_2X
 & \partial_1X\cdot\partial_2X \\ \partial_1X\cdot\partial_2X & -\partial_1X \cdot\partial_1X
\end{matrix}\right ) \, .
\end{equation}
Inserting this result into \eqn{FOFORM1} yields the first order form of the action
\begin{align}
S' = \int d^3\xi \, &\Bigr [ P_\mu\, \dot X^\mu + \frac{1}{2T\gamma^{00}}\, \Bigl \{ \,
P_\mu\, G^{\mu\nu}(X)\, \Bigl ( P_\nu  - \kappa \, C_{\nu\rho\kappa}(X)\, \{ X^\rho , X^\kappa\}\, \Bigr )
\nonumber \\ & +\frac{T^2}{2}\, \{ X^\mu, X^\nu\}\, \{ X^\rho, X^\kappa\}\, G_{\mu\rho}(X)\,
G_{\nu\kappa}(X)\, \Bigr \}\,
\Bigr ] ,
\label{S'}
\end{align}
with the usual definition of the Poisson bracket
$\{ X^\mu , X^\nu\} := \epsilon^{rs}\, \partial_rX^\mu\, \partial_sX^\nu$.
This formulation of the theory is a suitable starting point for a light-cone gauge.

\subsection{Gauge fixed action}

We now impose the light-cone gauge conditions
\begin{equation}
X^+= \tau , \qquad P_-=1 \, .
\label{lcg}
\end{equation}
Let us then specialize to the background of our inhomogenous pp-wave ansatz
\eqn{sugraansatz}
($r,s,t=1,\ldots, 6$, $M,N=1,\ldots, 9$)
\begin{align}
G_{++}&= \phantom{+} H(X^r), \quad G_{+-}=1, \quad G_{--}=0, \quad G_{MN}=\delta_{MN}=G^{MN},
 \quad C_{+rs}\neq 0 ,\nonumber \\
 G^{--}&= - H(X^r), \quad G^{+-}=1, \quad G^{++}=0 . \quad
\label{BG1}
\end{align}
The light-cone Hamiltonian $-P_+$ now follows from solving the equations of motions for $\gamma^{00}$ emerging
from \eqn{S'} for this specific background. Due to the gauge choices \eqn{lcg} many terms cancel
and one finds
\begin{equation}
{\cal H}_{LC} = -P_+ = \frac{1}{2}\, \Bigr ( P_M^2 -H(X^r) - \kappa\, C_{+rs}(X^t)  \,
\PB{X^r}{X^s}  +\frac{T^2}{2}\, (\PB{X^M}{X^N})^2\Bigl ) \, .
\end{equation}
Alternatively one may consider the gauge-fixed second order form of the
action which is obtained from \eqn{S'} upon reinserting
the solution of the equations of motion for the transverse momenta and the above
$P_+$. One then finds
\begin{equation}
{\cal L}_{GF}= \frac{1}{2}\,  \dot X_M^2 +\frac{1}{2}\, H(X^r) + \frac{\kappa}{2}\, C_{+rs}(X^t)\,
\PB{X^r}{X^s}  -\frac{T^2}{4}\, (\PB{X^M}{X^N})^2 \, .
\end{equation}
This constitutes the light-cone bosonic membrane action in the background \eqn{BG1}.

We now turn to the fermionic sector which has been neglected so far. The linear couplings to the
background fields are known for the complete
light-cone supermembrane from the supermembrane vertex operator construction of
\cite{Dasgupta:2000df}. From this we infer that next to the usual flat-space
fermion structure there is only one additonal Yukawa-type interaction term coupling to the three-form.
We then have (setting $T=\kappa=1$) the gauge fixed supermembrane lagrangian
\begin{align}
{\cal L}_{\text{APD}} &=
 \frac{1}{2}\, (D_\tau X_M)^2+ \frac{1}{2}\, H(X^r) + \frac{1}{2}\, C_{+rs}(X^t)\,
\PB{X^r}{X^s}  -\frac{1}{4}\, (\PB{X^M}{X^N})^2
 \nonumber \\
 & +i \,\theta^\dag \,D_\tau\theta  +i\,\theta^\dag\Gamma_M\,\{X^M,\theta\,\}
- \frac{i}{8}\, (\partial_r\,C_{+st})\, \theta^\dag \Gamma^{rst}\theta\, ,
\label{LGFsusy}
\end{align}
where $\theta$ are $SO(9)$ Majorana spinors. In \eqn{LGFsusy} we have also promoted
the $\tau$ derivatives to covariant ones
$D_\tau\, {\cal O}= \partial_\tau {\cal O}- \PB{\omega}{\cal O}$, where $\omega$ is the gauge field
of  area preserving diffeomorphisms (APD) whose equations of motion give rise to the remaining
`level-matching' contraint equations \eqn{Constraint}.
This formulation of the supermembrane allows for a $SU(N)$ matrix regularisation in the usual
fashion known from a flat space background \cite{de Wit:1988ig} modulo ordering ambiguities.

Next we go to a complex basis in the bosonic $SO(6)$ sector: $X^1,\ldots , X^6
\to Z^a, \Zb^{\bar a}$ with $a=1,2,3$ and $\bar a= \bar 1,\bar 2,\bar 3$  explicitly we take
\begin{align}
Z^1 &= \frac{1}{\sqrt{2}}\, (X^1+i\, X^2), \qquad
\Zb^{\bar{1}} = \frac{1}{\sqrt{2}}\, (X^1-i\, X^2)=(Z^1)^\dag, \qquad \text{etc.} \nn\\
X^1 &= \frac{1}{\sqrt{2}}\, (Z^1 +\Zb^{\bar 1}), \qquad\quad
X^2 = -\frac{i}{\sqrt{2}}\, (Z^1 -\Zb^{\bar 1}), \qquad \text{etc.}
\end{align}
and the metric reads, now letting $M=(a,\bar{a},i)$,
\be
\eta_{MN}=\left(\ba{ccc}
0~&~\delta_{a\bar{b}}~&~0\\
\delta_{\bar{b}a}~&~0~&~0\\
0~&~0~&~\delta_{ij}\ea\right)\,.
\ee
Note that depending on the context we employ a real or complex notation for the $SO(6)$ indices,
\textit{i.e.~}$M=(a,\bar a ,i)$ {\sl or} $M=(r,i)$, however maintaining the same symbol $X^M$
for the embedding coordinates\footnote{The embedding coordinates satisfies then the
following reality condition
$$ (X^M)^{\dagger}=\eta_{MN}X^{N}=X_{M}\qquad \text{where}\quad X^M=(Z^a,\bar Z^{\bar a},X^i)\,. $$}.
The distinction should be clear, however, from the context.

The background field data in this language from the supergravity analysis of
chapter 2 is
\begin{equation}
H(Z,\bar Z) = - \partial_aW(Z)\, \partial_{\bar a}\overline W(\Zb) \, , \qquad
C_{+ab} = \epsilon_{dab}\, \partial_{\bar d}\overline W(\Zb) \, , \qquad
C_{+\bar a\bar b} = \epsilon_{\bar d\bar a\bar b}\, \partial_{d}W(Z)
\, .
\end{equation}
In \eqn{LGFsusy} the $SO(9)$ gamma matrices $\Gamma^M$ are  $16\times 16$ nine-dimensional
(with the indices $M=a,\bar{a},i$)
and satisfy the Clifford algebra,
\be
\{\Gamma^{M},\Gamma^{N}\}=2\eta^{MN}\,.
\ee
With the charge conjugate matrix $C$,
\be
\ba{ll}
(\Gamma^{M})^{T}=(\Gamma_{M})^{\ast}=C\Gamma^{M}C^{-1}\,,~~~&~~C=C^{T}\,,
\ea
\ee
the 16-component spinor $\theta$ satisfies the Majorana condition
\be
\theta^{\dagger}=\theta^{T}C\,.
\label{Majorana}
\ee

We then have the following form of the gauge fixed supermembrane lagrangian
\begin{align}
{\cal L}_{\text{APD}} &=
 \frac{1}{2}\, D_\tau X^M\, D_\tau X_M -\frac{1}{4}\, \PB{X^M}{X^N}\PB{X_M}{X_N}
 +i \,\theta^\dag \,D_\tau\theta  + i\,\theta^\dag\Gamma_M\,\{X^M,\theta\,\} \nn\\
&
 -\frac{1}{2}\, \partial_aW\, \partial_{\bar a}\overline W + \frac{1}{2}\,
\epsilon_{dab}\, \partial_{\bar d}\overline W\,\PB{Z^a}{Z^b}
 + \frac{1}{2}\,   \epsilon_{\bar d\bar a\bar b}\, \partial_{d}W\,\PB{\bar Z^{\bar a}}{\bar Z^{\bar b}}
 \nonumber \\
 &-\frac{i}{8}\, \eps_{\bar d \bar a \bar b}
\, \partial_c\partial_d W\,  \theta^\dag\Gamma_{\bar c a b}\theta
 - \frac{i}{8}\, \eps_{ d  a  b}
\, \partial_{\bar c}\partial_{\bar d}  \overline W\,  \theta^\dag\Gamma_{c \bar a \bar b}\theta
\, .
\label{Lsupermem}
\end{align}
In the sequel we shall show that this action also arises from dimensional reduction
of ${\cal N}=1$ super Yang-Mills coupled to three chiral matter multiplets with a
superpotential dictated by $W(Z)$.
Before doing so we will present the matrix theory version of this supermembrane
theory and state the supersymmetry transformations.

\section{The $\cM$-theory matrix model description}

The standard matrix discretization procedure of the supermembrane action \eqn{Lsupermem}
 replaces the embedding coordinates by $N\times N$ matrices
and the Poisson-brackets by commutators, \textit{i.e}.
\be
X^M(\tau,\sigma_1\,\sigma_2) \longrightarrow (X^M)_{ij}(\tau)\,,\qquad
\PB{\,\cdot\,}{\,\cdot\,} \longrightarrow i \, [ \, \cdot\,  ,\,  \cdot\,  ]\,.
\ee
However, writing down the supersymmetric
matrix model associated to \eqn{Lsupermem} is nontrivial due to ordering
ambiguities of the matrices. Here we present our result first, and then
discuss the subtleties involved. We claim that the following
$\cM$-theory matrix model,
\be
\ba{ll}
\cL_{\text{MM}}=&\!\!\Tr\Big(\half D_{t}X^{M}D_{t}X_{M}+\quarter [X^{M},X^{N}][X_{M},X_{N}]+i\theta^{\dagger}
D_{t}\theta-
\theta^{\dagger}\Gamma^{M}[X_{M},\theta]\Big)\\
{}&{}\\
{}&+\textstyle{\frac{1}{2}}\,\Tr\Big(i\epsilon_{\bar{a}\bar{b}}{}^{c}[\bar{Z}^{\bar{a}},\bar{Z}^{\bar{b}}]\partial_{c}\, W+
i\epsilon_{ab}{}^{\bar{c}}[{Z}^{{a}},{Z}^{{b}}]\bar{\partial}_{\bar{c}}\,\overline{W}-
\partial_{a}
W\partial^{a}\overline{W}\Big)\\
{}&{}\\
{}&-\textstyle{\frac{i}{8}}\, \Tr\Big(\theta^{\dagger}\Gamma^{a}\Gamma^{\bar{1}\bar{2}\bar{3}}\Gamma^{b}\partial_{a}\Tr\left(\theta\partial_{b} W\right)
+\theta^{\dagger}\Gamma^{\bar{a}}\Gamma^{123}\Gamma^{\bar{b}}\bar{\partial}_{\bar{a}}\Tr\left(\theta
\bar{\partial}_{\bar{b}}
\overline{W}\right)\Big)\,, \label{MatrixLag}
\ea
\ee
enjoys four dynamical as well as four kinematical supersymmetries. Note that the Lagrangian above is determined by an arbitrary
holomorphic superpotential $W$ which is a scalar function of 3 hermitian matrices
$Z_a$. It goes over to the membrane lagrangian \eqn{Lsupermem} in the
$N\to \infty$ limit upon replacing commutators by Poisson brackets. While this is rather
obvious for the first two lines of \eqn{MatrixLag}, it is not much so for the Yukawa-type
 terms of the
last line of \eqn{Lsupermem}. In order to make the comparison we note that,
upon ignoring
the ordering of the matrix valued
$\theta$ and $Z$ fields, we have
\begin{align}
\Tr\Big(\theta^{\dagger}\Gamma^{a}\Gamma^{\bar{1}\bar{2}\bar{3}}\Gamma^{b}\partial_{a}\Tr\left(\theta
\partial_{b} W \right)\Big)&
    \rightarrow  \partial_e \partial_d W\, \theta^{\dagger}\Gamma^{e}
    \Gamma^{\bar 1 \bar 2 \bar 3} \Gamma^{d} \theta
=   \partial_e \partial_d W\,  \theta^{\dagger}\Gamma^{e}
        (\textstyle{\frac{1}{6} }\epsilon_{\bar a \bar b \bar c} \Gamma^{\bar a \bar b \bar c})\Gamma^{d} \theta \nn\\
&=   \epsilon_{\bar d \bar a \bar b} \partial_d \partial_c W \, \theta^\dagger
\Gamma_{\bar c a b} \theta
\end{align}
matching the corresponding terms in \eqn{Lsupermem}.

$W$ is an arbitrary $\mbox{U}(N)$ singlet and holomorphic in $Z_a$, and
$\overline{W}$ is the complex conjugate $W^{\dagger}$.
More explicitly, $W$ is a function of traces, like $\Tr(Z^{a_{1}}Z^{a_{2}}\cdots Z^{a_{n}})$,
so we allow multi-traces, for instance.
$\partial_{a}W$ is  matrix-valued as we suppress the matrix indices in our presentation.
The time derivative is the gauge covariant one such as  $D_{t}X^{i}=\frac{{\rmd}~}{\rmd t} -i [A_{0},X^{i}]$, $A_0$
is the matrix field corresponding to the APD gauge field $\omega$ in  \eqn{Lsupermem}.

The $4+4$ supersymmetries  are realized as
\be
\ba{ll}
\delta A_{0}=i\theta^{\dagger}\varepsilon\,,~~~&~~~
\delta X^{M}=i\theta^{\dagger}\Gamma^{M}\varepsilon\,,\\
{}&{}\\
\multicolumn{2}{c}{
\delta\theta=\frac{1}{2}\,\Big(\Gamma^{M}D_{t}X_{M}-\frac{i}{2}\,\Gamma^{MN}[X_{M},X_{N}]+
\frac{1}{4}\,\Gamma^{a}\Gamma^{\bar{1}\bar{2}\bar{3}}
\partial_{a}W+\frac{1}{4}\Gamma^{\bar{a}}\Gamma^{123}\bar{\partial}_{\bar{a}}\overline{W}\Big)\varepsilon +\eta\,.}
\ea
\ee
The supersymmetry parameters $\varepsilon$ and $\eta$ are Majorana spinors
 $\varepsilon^{\dagger}=\varepsilon^{T}C$, $\eta^{\dagger}=\eta^{T}C$ and satisfy the following projection property
\be \ba{ll} \varepsilon=P\varepsilon\,,\qquad
\eta=P\,\eta\,,~~~&~~~P=\textstyle{\frac{1}{8}}(
\Gamma^{123}\Gamma^{\bar{3}\bar{2}\bar{1}}+\Gamma^{\bar{3}\bar{2}\bar{1}}\Gamma^{123})\,.
\ea \ee Essentially the projector $P$ leaves only the $\mbox{SU}(3)$
singlet sector,\footnote{ $P$ decomposes further into two orthogonal
  projections, $P=P_{+}+P_{-}$ where
  $P_{+}=\textstyle{\frac{1}{8}}\Gamma^{123}\Gamma^{\bar{3}\bar{2}\bar{1}}$,
  $P_{-}=\textstyle{\frac{1}{8}}\Gamma^{\bar{3}\bar{2}\bar{1}}\Gamma^{123}$
  satisfying $P_{\pm}=P_{\pm}^{2}=P_{\pm}^{\dagger}$,
  $P_{+}P_{-}=P_{-}P_{+}=0$ and $\tr P_{\pm}=2$. We furthermore note the identities
$\Gamma^{\bar 1 \bar 2\bar 3}\, \Gamma^a\, P=0$ and $\Gamma^{123}\, \Gamma^{\bar a}\, P=0$.} which has four
nontrivial components, since $P=P^{2}=P^{\dagger}$ and $\tr P=4$.  We
thus have four dynamical supersymmetries parametrzed by $\varepsilon$ and four
kinematical supersymmetries parametrizes $\eta$ intact, matching the supergravity
picture.

In order to verify the supersymmetry invariance it is noteworthy that
the following terms, cubic in $\theta$, vanish identically
\be
\Tr\!\left[\theta^{\dagger}\Gamma^{a}\varepsilon\partial_{a}\Tr\!\left(
\theta^{\dagger}\Gamma^{b}\Gamma^{\bar{1}\bar{2}\bar{3}}\Gamma^{c}\partial_{b}\Tr
(\theta
\partial_{c}W)\right)\right]=0\,,
\ee
since it essentially corresponds to anti-symmetrizing three of two-component spinor indices. Other useful identities are\footnote{This can be checked by noting
\[
[Z^{a},\partial_{a}\Tr(Z^{b_{1}}Z^{b_{2}}\cdots Z^{b_{n}})]=\sum_{l=1}^{n} \left(
Z^{b_{l}}\cdots Z^{b_{n}}Z^{b_{1}}\cdots Z^{b_{l-1}}-Z^{b_{l+1}}\cdots Z^{b_{n}}Z^{b_{1}}\cdots Z^{b_{l}}\right)=0\,.
\]An analogue identity holds for the holomorphic superpotential in the supermembrane action in terms of  Poisson bracket, $\{Z^a,\partial_{a}W(Z)\}=0$.}
\be
\ba{ll}
[Z^{a},\partial_{a}W]=0\,,~~~&~~~[\bar{Z}^{\bar{a}},\bar{\partial}_{\bar{a}}\overline{W}]=0\,.
\ea
\ee
Having established the supermembrane and matrix theory description of our $\cM$-theory pp-wave
background, we now proceed to investigate how these actions can be reexpressed as dimensional
reduction of four-dimensional ${\cal N}=1$ supersymmetric Yang-Mills theory coupled to three
chiral matter supermultiplets.


\section{The ${\cal N}=1$ supersymmetry  description}
Our discussions so far make it evident that the supermembrane or  matrix
theory can be
alternatively understood as the dimensional reduction of ${\cal N}=1, D=4$
super Yang-Mills theory to one dimension. In this section we start from the supermembrane action
and rephrase it in a way where ${\cal N}=1$ symmetry is more manifest.
We first show that in the bosonic sector the interactions are correctly
given by the F-term and D-term potentials from the true superpotential which
includes the familiar cubic superpotential of ${\cal N}=4$ theory, in addition to
the superpotential of the supergravity background. We then proceed to decompose the
SO(9) Majorana spinor into 4 copies of 2-component Weyl spinors, and in particular
check that the Yukawa couplings are determined by the superpotential, just as one
would expect from ${\cal N}=1$ supersymmetry.
\subsection{The  superpotential}

The bosonic part of the gauge-fixed supermembrane or gauge theory of area-preserving diffeomorphism
Lagrangian reads \eqn{LGFsusy}
\begin{align}
{\cal L}_{GF, bos}&= \frac{1}{2}\, (D_\tau X_M)^2 - V(X), \nonumber\\
\text{with}\quad
V(X)&=-\frac{1}{2}\, H(X^r) - \frac{1}{2}\, C_{+rs}\, \PB{X^r}{X^s} + \frac{1}{4}\,
 (\PB{X^M}{X^N})^2 \, . \label{Vx}
\end{align}
We note the decomposition of the last piece of the scalar potential in the pure $SO(6)$
sector ($r,s=1,\ldots, 6$)
\begin{align}
\frac{1}{4}\,\int d^2\sigma\, (\PB{X^{r}}{X^{s}})^2 &=
 \int d^2\sigma\, \Bigl (\, \PB{Z^a}{Z^b}\,  \PB{\Zb^{\bar a}}{\Zb^{\bar b}}
- \frac{1}{2}\, (\PB{Z^a}{\Zb^{\bar a}})^2 \Bigr )  \nonumber\\
 &=
 \int d^2\sigma\, \Bigl (\, \frac{1}{2}\, \epsilon_{dab}\, \PB{Z^a}{Z^b}\,
\epsilon_{\bar d\bar a\bar b}\,  \PB{\Zb^{\bar a}}{\Zb^{\bar b}}
- \frac{1}{2}\, (\PB{Z^a}{\Zb^{\bar a}})^2 \Bigr ) \, ,
\end{align}
where use of Jacobi's identity for the Poisson brackets has been made.
With the help of this we can now rewrite the scalar potential $V(X^i,Z^a,Z^{\bar a})$ into
F-term and D-term pieces along with the SO(3) and SO(6) mixed contributions ($m=1,2,3$):
\begin{align}
V(X,Z,\Zb) &= \frac{1}{2}\, \Bigl (\,\partial_dW - \epsilon_{dab}\, \PB{Z^a}{Z^b}\, \Bigr)
\,\Bigl (\,\partial_{\bar d}\overline W - \epsilon_{\bar d\bar a\bar b}\,
\PB{\Zb^{\bar a}}{\Zb^{\bar b}}\, \Bigr )
 -\frac{1}{2}\, (\, \PB{Z^a}{\Zb^{\bar a}}\, )^2 \nonumber \\
& +\frac{1}{4}\, (\PB{X^i}{X^j})^2 + \PB{X^i}{Z^a}\, \PB{X^i}{\Zb^{\bar a}}\, .
\end{align}
Hence the F-term piece of the potential is governed by the holomorphic superpotential
\begin{align}
{\cal W}(Z^1,Z^2,Z^3) = \int d^2\sigma\, \Bigl (\, W(Z^1,Z^2,Z^3) -\frac{1}{3}\, \epsilon_{abc}\,
Z^a\, \PB{Z^b}{Z^c}\, \Bigr )\, .
\end{align}

\subsection{The fermionic terms }

We now turn to a rewriting of the $\grSO(9)$ spinors in \eqn{Lsupermem}
or respectively \eqn{MatrixLag} in an $\grSO(3)\times \grSO(6)$
split following the conventions of~\cite{Kim:2003rza}, appendix A. For this we decompose the Dirac matrices according to
($i=1,2,3, r=1,\ldots,6$)
\begin{align} \label{eqn:SO9gammas}
\Gamma_i & =
  \left( \begin{matrix}
  -\sigma_i \otimes \unit_4  &  0 \\
  0                             &  \sigma_i \otimes \unit_4
  \end{matrix} \right) \, ,\quad
\Gamma_r  =
  \left( \begin{matrix}
  0                              &  \unit_2 \otimes \rho_r \\
  \unit_2 \otimes \rho_r^\dag  &  0
  \end{matrix} \right) \, ,
\end{align}
where $\sigma_i$ are the three Pauli matrices and the $4\times 4$ matrices $\rho_r$ and
$\rho_r^\dag$ satisfy
\be
\rho_r\,\rho^\dag_s+\rho_s\,\rho^\dag_r=\rho_r^\dag\,\rho_s+\rho_s^\dag\,\rho_r=2\, \delta_{rs}\,\unit_4\, .
\ee
The charge conjugation matrix in this representation is given by
\be C_9 =
  \left( \begin{matrix}
  0                          &  \epsilon^{\alpha \beta} \otimes \unit_4 \\
  -\epsilon_{\dot \alpha \dot \beta} \otimes \unit_4  &  0
  \end{matrix} \right) \; ,
\ee allowing one to write the spinor as\footnote{We use the standard index free Weyl spinor notation with the
convention: $\lambda\psi:=-\eps^{\alpha\beta}\lambda_\alpha\psi_\beta=\psi\lambda$ and
$\bar\lambda\bar\psi:=\eps^{\dot\alpha\dot\beta}\bar\lambda_{\dot\alpha}\bar\psi_{\dot\beta}=\bar\psi\bar\lambda$. Moreover
$i(\sigma^2)^{\alpha\beta}=\epsilon^{\alpha\beta}$ and
$(\lambda_\alpha)^\ast=\bar\lambda_{\dot\alpha}$.}
 \be \theta =  { \theta_{\alpha A} \choose \bar\theta^{\dot\alpha}_A } \; , \theta^\dagger =
\theta^T\, C_9 = ( \bar\theta_{\dot\alpha A} ~~ \theta^{ \alpha}_A)\, ,
 \qquad \alpha=1,2\:,\;\; A=1,\ldots,4 \; , \label{realspinors}
\ee
where $\theta_{\alpha A}$ and $\bar\theta_{\dot\alpha\, A}$ are now four 2-component
Weyl spinors respectively. In terms of these the
Yukawa couplings to the $\grSO(6)$ scalars $X^r$ may be reexpressed as
\begin{align}
\label{Yukawas}
\PB{\theta^\dag\Gamma_r}{\theta}\, X^r &= \PB{\theta_A}{\theta_B}\, (\rho^\dag_r)_{AB}\, X^r
+ \PB{\bar\theta_A}{\bar\theta_B}\,
 (\rho_r)_{AB}\, X^r=\\
\PB{\theta_{A}}{\theta_{ B}}\,  &\, \Bigl [ (\Omega_a)_{AB}\, Z^a +
(\Omega_{\bar a})_{AB}\, \Zb^{\bar a} \Bigr ]
+\PB{\bar\theta_{A}}{\bar\theta_{ B}}\,
\Bigl [ (\bOmega_a)_{AB}\, Z^a +
(\bOmega_{\bar a})_{AB}\, \Zb^{\bar a} \Bigr ] \, ,\nn
\end{align}
where we have introduced the $4\times 4$ matrices
$\Omega_a$ and  $\Omega_{\bar a}$ ($a=1,2,3; \bar a=\bar 1,\bar 2,\bar 3$) via
\begin{align}
\Omega_1&=\frac{1}{\sqrt{2}}\, (\rho_1 -i\rho_2)\, , \qquad
\Omega_{\bar 1}=\frac{1}{\sqrt{2}}\, (\rho_1 +i\rho_2)\, , \qquad\text{etc.} \nn\\
\bOmega_1&=\frac{1}{\sqrt{2}}\, (\rho_1^\dag -i\rho_2^\dag)\, , \qquad
\bOmega_{\bar 1}=\frac{1}{\sqrt{2}}\, (\rho_1^\dag +i\rho_2^\dag)\, , \qquad\text{etc.}
\end{align}
which satisfy
\begin{align}
  \Omega_{\bar a} \bOmega_b +  \Omega_{b} \bOmega_{\bar a} =
  \bOmega_{\bar a} \Omega_b +  \bOmega_{b} \Omega_{\bar a} &= 2\, \eta_{\bar{a} b}\, \unit_4 \, ,\nn\\
 \Omega_{a} \bOmega_b +  \Omega_{b} \bOmega_{a} =
  \bOmega_{a} \Omega_b +  \bOmega_{b} \Omega_{a} &= 0 \, ,\nn\\
 \Omega_{\bar a} \bOmega_{\bar b} +  \Omega_{\bar b} \bOmega_{\bar a} =
  \bOmega_{\bar a} \Omega_{\bar b} +  \bOmega_{\bar b} \Omega_{\bar a} &= 0 \, .
\end{align}
It is useful to employ a definite representation for the antisymmetric $\rho_r$ matrices:
\begin{align}
\rho_1&= -\unit \otimes i\sigma_2\, , \qquad \rho_2=-\sigma_3\otimes\sigma_2 \, ,\nn\\
\rho_3&= -i\sigma_2 \otimes \sigma_3\, ,\qquad \rho_4=-\sigma_2\otimes\unit \, ,\nn\\
\rho_5&= -i\sigma_2 \otimes \sigma_1\, , \qquad \rho_6=-\sigma_1\otimes\sigma_2 \, .
\end{align}
In terms of these one finds
\be
\Omega_a\, Z^a = \sqrt{2}\, \left (\begin{matrix}  0 & 0 &0 & 0\cr 0 & 0& -Z^3 & Z^2 \cr
0 & Z^3 & 0 & -Z^1 \cr 0 & -Z^2 & Z ^1 & 0\end{matrix} \right )\, , \qquad
\Omega_{\bar a}\, \Zb^{\bar a} = \sqrt{2}\,
\left (\begin{matrix}  0 & -\Zb^{\bar 1} & -\Zb^{\bar 2} & - \Zb^{\bar 3}\cr
 \Zb^{\bar 1} & 0& 0 & 0 \cr
\Zb^{\bar 2} & 0 & 0 & 0 \cr  \Zb^{\bar 3} & 0 & 0 & 0\end{matrix} \right ) \, .
\ee
We are led to identify $\theta_0$ with the ${\cal N}=1$ gluino $\lambda$ and the remaining components with
the $\grSU(3)$ matter fermions $\psi^a$ and $\bar \psi^{\bar a}$
in the ${\bf 3}$ and ${\bf \bar 3}$ representations via
\be
\theta_A= (\lambda, \psi^1,\psi^2,\psi^3) \qquad
\bar\theta_A= (\bar\lambda, \bar\psi^{\bar 1},\bar\psi^{\bar 2},\bar\psi^{\bar 3}) \, .
\ee
This leads us to the compact expressions
\begin{align}
\PB{\theta_A}{\theta_B} (\Omega_a)_{AB}\, Z^a &= -\sqrt{2}\, \eps_{abc}\, Z^a\, \PB{\psi^b}{\psi^c}
\, , \nn\\
\PB{\theta_A}{\theta_B} (\Omega_{\bar a})_{AB}\, \Zb^{\bar a} &= -2\,\sqrt{2}\, \Zb^{\bar a}\,
\PB{\lambda}{\psi^a} \nn\\
\PB{\bar\theta_A}{\bar\theta_B} (\bOmega_a)_{AB}\, Z^a &= 2\,\sqrt{2}\, Z^{ a}\,
\PB{\bar\lambda}{\bar\psi^{\bar a}}
\, , \nn\\
\PB{\bar\theta_A}{\bar\theta_B} (\bOmega_{\bar a})_{AB}\, \Zb^{\bar a} &=
\sqrt{2}\, \eps_{\bar a\bar b\bar c}\, \Zb^{\bar a}\, \PB{\bar\psi^{\bar b}}{\bar\psi^{\bar c}}\, ,
\end{align}
again suppressing the 2-component Weyl spinor indices. Upon inserting this into \eqn{Yukawas} reproduces the
Yukawa couplings and part of the superpotential couplings of
${\cal N}=1$ super Yang-Mills coupled to 3 chiral multiplets in the $\grSU(3)$.

The remaining fermion term coupling to the three-form $C_{+rs}$ of \eqn{LGFsusy} (compare \eqn{Lsupermem}) reads
\be
\label{Wterm}
\frac{i}{8}\,
 \partial_r C_{+st}\, \theta^\dag\Gamma^{rst}\theta = \frac{i}{8}\, \eps_{\bar d \bar a \bar b}
\, \partial_c\partial_d W\, \theta^\dag\Gamma_{\bar c a b}\theta
+\frac{i}{8}\, \eps_{ d a b} \, \partial_{\bar c}\partial_{\bar d}
\overline W\, \theta^\dag\Gamma_{c \bar a \bar b}\theta \, . \ee
In the $\grSO(3)\times \grSO(6)$ split the three-index Dirac matrices
$\Gamma_{\bar c a b}$ take the form
\be \Gamma_{\bar c a b} = \left
  ( \begin{matrix} 0 &\unit_2\otimes \Omega_{\bar c a b} \cr
    \unit_2\otimes \bOmega_{\bar c a b} & 0 \cr
\end{matrix} \right )\, , \quad \theta^\dag\Gamma_{\bar c a b}\theta =
\theta_A\theta_B\, (\Omega_{\bar c a b})_{AB} +
\bar\theta_A\bar\theta_B\, (\bOmega_{\bar c a b})_{AB}\, , \ee with
$\Omega_{\bar c a b} := \Omega_{[\bar c}\, \bOmega_{a}\, \Omega_{b]}$
and $\bOmega_{\bar c a b} := \bOmega_{[\bar c}\, \Omega_{a}\,
\bOmega_{b]}$ antisymmetrized with unit weight. One then shows using
the above representation that
\begin{align}
\theta_A\theta_B\, (\Omega_{\bar c a b})_{AB} &= 2\sqrt{2}\, \psi^c\psi^d\, \eps_{abd} \nn\\
\bar\theta_A\bar\theta_B\, (\bOmega_{\bar c a b})_{AB} &= -2\sqrt{2}\, (\, \eta_{\bar c a}\,
\bar\lambda\bar\psi^{\bar b} -\eta_{\bar c b}\, \bar\lambda\bar\psi^{\bar a} \,)
\end{align}
Hence we have in \eqn{Wterm} \be \frac{i}{8}\, \eps_{\bar d \bar a
  \bar b} \, \partial_c\partial_d W\, \theta_A\theta_B\, (\Omega_{\bar
  c a b})_{AB} = \frac{i}{\sqrt{2}} \, \partial_c\partial_d W\,
\psi^c\psi^d\, , \ee the expected fermionic coupling in the matter
sector to the holomorphic superpotential $W(Z^a)$, whereas the
nonholomorphic second term in \eqn{Wterm} drops out as it should: \be
\frac{i}{8}\, \eps_{\bar d \bar a \bar b} \, \partial_c\partial_d W\,
\bar\theta_A\bar\theta_B\, (\bOmega_{\bar c a b})_{AB} =
\frac{i}{\sqrt{2}}\, \eps_{\bar d\bar c \bar b}\, \partial_c\partial_d
W\, \bar\lambda\bar\psi^{\bar b}
=0\,. \label{nonholomorphic_second_term_with_W}\ee and the analogous
terms for the hermitian conjugate contributions.

Upon collecting everything we indeed find an ${\cal N}=1$ super Yang-Mills model of area preserving
diffeomorphisms coupled to three chiral multiplets transforming in the fundamental representation
of $\grSU(3)$ dimensionally reduced to one-time dimension:
\be
\ba{ll}
{\cal L}_{{GF},{~susy}} =&\textstyle{\frac{1}{2} (D_\tau X^i)^2 -\frac{1}{4}\, (\PB{X^i}{X^j})^2
+ D_\tau Z^{a}\, D_\tau \Zb^{\bar a}- \PB{X^i}{Z^a}\, \PB{X^i}{\Zb^{\bar a}}}\\
{}&{}\\
{}&\textstyle{ - \frac{1}{2}\partial_{a}{\cal W}(Z)\bar{\partial}_{\bar{a}}\overline{\cal W}(\Zb)
 +\frac{1}{2}\,  \PB{Z^a}{\Zb^{\bar a}}^2}\\
 {}&{}\\
{}& +2i \lambda\, D_\tau{\bar\lambda}  +2i\, \lambda\sigma_i\,\PB{X^i}{\bar\lambda}
+2i \psi^a\, D_\tau{\bar\psi^{\bar a}} +2i\, \psi^a\sigma_i\,\PB{X^i}{\bar\psi^{\bar a} }\\ {}&{}\\
   {} & -i\,2\sqrt{2}\, \Zb^{\bar a}\, \PB{\lambda}{\psi^a}
 +i\, 2\sqrt{2}\, Z^{a}\, \PB{\bar\lambda}{\bar\psi^{\bar a}}\\
 {}&{}\\
 {}&\textstyle{ - \frac{i}{\sqrt{2}}}\psi^a\partial_{a}{\dis{\int}d^2 \sigma'}\,
\psi^{b}(\sigma')\frac{\partial}{\partial Z^b(\sigma')}{\cal W}(Z)
 + \frac{i}{\sqrt{2}} \bar\psi^{\bar a}\bar{\partial}_{\bar{a}}{\dis{\int}d^2 \sigma'}\,
 \psi^{\bar b}(\sigma')\frac{\bar\partial}{\bar{\partial} {\bar Z}^{\bar{b}}(\sigma')} \overline{\cal W}(\Zb)\,.
\ea
\label{GFsusymembrane}
\ee
Here the holomorphic superpotential is given by an integral over the two-dimensional space
like components of the membrane worldsheet
\begin{align}
\label{superpot}
{\cal W}(Z^a) = \int d^2\sigma\, \Bigl (\, W(Z^a) -\frac{1}{3}\, \epsilon_{abc}\, Z^a\,
\PB{Z^b}{Z^c}\, \Bigr )\,,
\end{align}
and any derivative acting on it, as $\partial_{a}{\cal W}$, must be understood as a
functional derivative with respect to $Z^a(\sigma)$.

Now we are ready to obtain the Matrix theory action utilizing again the familiar
discretisation procedure of replacing the Poisson brackets with matrix commutators:
\be
\ba{ll}
\cL =\Tr \Big[\!\!\!\!& {1\over 2}D_t X^i D_t X^i + {1\over 4} [X^i, X^j]^2
            + D_t Z^a D_t \bar{Z}^{\bar a}  + [X^i,Z^a][X^i,\bar{Z}^{\bar a}]
\\{}&{}\\
 & \textstyle{
-\frac{1}{2}\partial_a{\cal W}(Z)\bar{\partial}_{\bar{a}}\overline{\cal W}(\bar{Z})
    -{1\over 2}[Z^a, \bar{Z}^{\bar a}]^2}
 + 2i \lambda D_t \bar \lambda - 2\lambda \sigma^i[X_i , \bar\lambda]
 \\ {}&{}\\{}&
      + 2i \psi^a D_t \bar \psi^{\bar a} - 2\psi^a \sigma^i [X_i, \bar\psi^{\bar a}]
 + 2 \sqrt 2 \bar{Z}^{\bar a} [\lambda, \psi^a] -  2\sqrt 2 Z^a [\bar \lambda, \bar \psi^{\bar
    a}]
\\{}&{}\\
{}&\textstyle{ - \frac{i}{\sqrt{2}}}\psi^a\partial_{a}{\dis{\Tr}}\left(\psi^{b}\partial_{b}{\cal W}(Z)\right)
 + \frac{i}{\sqrt{2}} \bar\psi^{\bar a}\bar{\partial}_{\bar{a}}{\dis{\Tr}}\left(
 \psi^{\bar b}\bar{\partial}_{\bar{b}} \overline{\cal W}(\Zb)\right)    \Big]\,.
       \label{MatrixLag2spinors}
\ea \ee

This constitutes our main result: $\cM$-theory in a generalized
pp-wave background with gauge flux described by a holomorphic function
$W(Z^a)$ has its  Matrix theory description as the dimensional
reduction of ${\cal N}=1$, $D=4$, $U(N)$ Yang-Mills
theory coupled to three chiral supermultiplets with
superpotential $\cW(Z)$ given by
\be
{\cal W} (Z) = W(Z) - \frac{i}{3} \epsilon_{abc} \Tr\!\left( Z^a [ Z^b , Z^c ]\right)  .
\ee
In closing we would like to remark that the membrane or matrix theory actions
found in the above are ${\cal N}=1$ supersymmetric irrespective
of the form of $W(Z)$. However, different orderings of the fields appearing
in $W(Z)$ lead to distinct
marix models, but are equivalent in supergravity or supermembrane theory. In addition
functions $W(Z)$ are possible containing Poisson-brackets or commutators of the
holomorphic fields $Z^a$, which
would have to be understood as being of non-geometric origin.

\section{Discussions}
In this paper we have considered a class of supersymmetric pp-wave solutions
in 11 dimensional supergravity.
The holomorphic function which describes the
configuration is  related to the superpotential of the Yang-Mills quantum mechanics
which comes from the discretized supermembrane action in the relevant background.

Following the spirit of \cite{Banks:1996vh}, it is natural
to conjecture that the Yang-Mills
quantum mechanics we have derived in this paper should provide
 a Matrix theory description of M-theory in the inhomogeneous
pp-wave backgrounds. As an alternative to the supermembrane action,
Matrix theory can be also obtained
as discrete lightcone quantization (DLCQ) of M-theory \cite{Susskind:1997cw}. 
In practice, one compactifies
M-theory on a small circle and at the same time performs an infinite boost.
The quantized lightlike
momentum $N$ is translated into the number of D0-branes through T-duality, and
the large $N$ Yang-Mills quantum mechanics of D0-branes gives the Matrix theory.

The generalization of DLCQ prescription of M-theory to nontrivial curved background was 
considered for
instance in \cite{Taylor:1998tv,Taylor:1999gq}, 
where the authors studied low energy dynamics of D0-branes in weakly curved backgrounds. 
As a simple but nontrivial example, one can
consider the maximally supersymmetric 11 dimensional plane-wave and perform DLCQ of
M-theory \cite{Dasgupta:2002hx}. Although the scalar curvature of 11 dimensional background
vanishes, the IIA configuration becomes singular when $H=G_{++}\rightarrow -4$.
In \cite{Dasgupta:2002hx} it is verified that in the small $H$ approximation the D0-brane
dynamics in the weakly curved background limit coincides with the regularized
supermembrane action, or the BMN matrix theory \cite{Berenstein:2002jq}.

One can also apply the method of \cite{Taylor:1999gq} to those solutions we studied in 
this paper. First of all, one can easily verify that $H=G_{++}$ again translates into the scalar
potential of the Yang-Mills theory, with Tseytlin's symmetrized trace prescription 
\cite{Tseytlin:1997csa} for matrix fields. This way we can resolve the ambiguity of 
matrix ordering problem utilizing the microscopic description through open string
excitations. The rest of the action should agree with the supermembrane prescription, 
since the various terms are related by $\cN=1$ supersymmetry. Summarizing, although 
DLCQ description for generic pp-waves has a drawback of limited validity due to the
singularity of IIA background, it in principle can fix the ordering problem of Matrix
regularization we encounter in supermembrane action. It will be certainly interesting
to further explore D0-brane dynamics in the T-dual background in IIA supergravity. 

For the solutions we studied in this paper, 
the flux is turned on along the six dimensional subspace only, and
the isometry group contains $SO(3)$. It is thus natural to view the matrix model
as originating from a four dimensional field theory. One might ask whether it is
possible to turn on a constant flux on $\mathbb{R}^3$, without breaking $SO(3)$.
This is exactly the mass deformation which transforms the ordinary Matrix theory
\cite{Banks:1996vh} into the BMN matrix model \cite{Berenstein:2002jq}, and
as a result there will be a cubic interaction term $\Tr X^1[X^2,X^3]$ in the matrix
model.

It is elucidated in \cite{Kim:2003rza} that, as a four dimensional super Yang-Mills theory,
the mass parameter
is related to the choice of putting the field theory on $\mathbb{R}\times S^3$
instead of $\mathbb{R}^{1,3}$. Since this freedom relies on classical superconformal
invariance, we expect it is not possible in general to have a constant flux in
$\mathbb{R}^3$, unless $W$ is cubic in $Z$. Among these the most interesting is
probably the so-called $\beta$-deformation which is known to be exactly marginal
as a 4 dimensional quantum field theory \cite{Lunin:2005jy}.
By $\beta$-deformation, the matrix commutator is replaced by
\be
[X , Y ] = X Y - Y X \longrightarrow e^{i\beta} X Y - e^{-i\beta} Y X ,
\ee
for a constant $\beta$. In the context of Matrix theory, this deformation
is considered in \cite{shimada} and the stable membrane solutions of different topology
are studied in the continuum limit.

It will be certainly very interesting to consider
BPS objects with different dimensions in the matrix models described in this paper.
For BMN matrix model, readers are referred to {\it e.g.} \cite{Bak:2002rq,Park:2002cba,Kim:2002tj}
for the study of BPS configurations.


\subsection*{Acknowledgements}

We would like to thank H. Shimada for crucial discussions on the existence
of 11 dimensional supersymmetric pp-waves with non-constant flux.
The work of J.P.~is supported by the Volkswagen
Foundation. NK would like to thank Humboldt University and the SFB 647 `Space-Time-Matter'
for their hospitality during the initial stages of this work.
The research of JHP is supported in part by the Korea Science and Engineering Foundation grant funded by the Korea government (R01-2007-000-20062-0), and the research of JWK, NK and JHP is supported
by the Center for Quantum Spacetime of Sogang University with grant number R11 - 2005 - 021.
The research of NK is also partly supported by the Korea Research Foundation Grant,
No. KRF-2007-314-C00056 and No. KRF-2007-331-C00072.

\end{document}